# Diagnosing syndromes of biosphere-atmosphere-socioeconomic change

Wantong Li,[1,2,3,*] Gregory Duveiller,[1] Fabian Gans,[1] Jeroen Smits,[4]
Guido Kraemer,[5] Dorothea Frank,[1] Miguel D. Mahecha,[5,6] Ulrich Weber,[1]
Mirco Migliavacca,[7] Andrej Ceglar,[8] Trevor F. Keenan,[2,3] Markus Reichstein[1,9,**]

[1]Max Planck Institute for Biogeochemistry, Biogeochemical Integration, Germany
[2]Department of Environmental Science, Policy and Management, UC Berkeley, Berkeley, CA, USA
[3]Climate and Ecosystem Sciences Division, Lawrence Berkeley National Laboratory, Berkeley, CA, USA
[4]Global Data Lab, Institute for Management Research, Radboud University, Netherlands
[5]Remote Sensing Center for Earth System Research, Leipzig University, Leipzig, Germany
[6]German Centre for Integrative Biodiversity Research (iDiv), Halle-Jena-Leipzig, Germany
[7]European Commission, Joint Research Centre (JRC), Ispra, Italy.
[8]Climate Change Centre of the European Central Bank, Frankfurt Am Main, Germany
[9]ELLIS Unit Jena, Germany
*Correspondence: wantongli@berkeley.edu
**Correspondence: mreichstein@bgc-jena.mpg.de

## SUMMARY

It is increasingly recognized that the multiple and systemic impacts of Earth system change threaten the prosperity of society through altered land carbon dynamics, freshwater variability, biodiversity loss, and climate extremes. For example, in 2022, there are about 400 climate extremes and natural hazards worldwide, resulting in significant losses of lives and economic damage[1]. Beyond these losses, comprehensive assessment on societal well-being, ecosystem services, and carbon dynamics are often understudied. The rapid expansion of geospatial, atmospheric, and socioeconomic data provides an unprecedented opportunity to develop systemic indices to account for a more comprehensive spectrum of Earth system change risks and to assess their socioeconomic impacts. We propose a novel approach based on the concept of syndromes that can integrate synchronized changes in biosphere, atmosphere, and socioeconomic trajectories into distinct co-evolving phenomena. While the syndrome concept was applied in policy related to environmental conservation, it has not been deciphered from systematic data-driven approaches capable of providing a more comprehensive diagnosis of anthropogenic impacts. By advocating interactive dimensionality reduction approaches, we can identify key interconnected socio-environmental changes as syndromes from big data. We recommend future research tailoring syndromes by incorporating granular data, particularly socio-economic, into dimensionality reduction at different spatio-temporal scales to better diagnose regional-to-global atmospheric and environmental changes that are relevant for socioeconomic changes.

## INTRODUCTION

Since the beginning of the 21[st] Century, increasing attention has been given to assessing and mitigating the risks of global change, while achieving sustainable development goals (SDGs). Climate scenario analyses and integrated assessment modeling can guide SDG policies[2]. Different temperature trajectories, coming from carbon emission pathways, give powerful (though simplified) narratives of climate change which motivate initiatives of carbon neutrality[3]. Planetary boundary frameworks warn of critical thresholds of global change due to anthropogenic factors, measured by the stability of the Earth system relative to Holocene levels[4,5]. Evaluations of physical, biological and biogeochemical processes, including nutrient cycles, land water, biodiversity, natural ecosystems, climate, and aerosols, indicate that seven of eight key

Earth system boundaries have been exceeded. Theories of resilience and complex systems further aid in diagnosing the stability of socio-ecological change[6].

Over the past three to four decades, climate change has destabilized terrestrial ecosystems and their associated net land carbon uptake[7]. This has been coupled with critical regional changes, such as those in the Amazon forest system, that are hypothesized to be approaching a tipping point [8]. Prognostic analyses of these changes rely heavily on modeling which account for a broad, though incomplete, range of socio-economic and natural processes. However, such modeling remains challenging due to its strong dependence on model structure and parameterization, which often results to significant uncertainty in projections[9,10].

The atmosphere, the biosphere, and the socioeconomic sphere have changed rapidly during the Anthropocene, necessitating comprehensive diagnostics to track and understand these shifts. While observations span scales from individual sites to satellite monitoring, and from counties to the global scale, only a subset of processes can be effectively captured through abundant observations. Moreover, only a subset of these observations has been used to constrain model simulations. Importantly, although there exists a broad array of indices for tracking change, a unifying framework is lacking. Examples of a broad array of indices can be that Steffen et al. 2004 & 2015[11,12] established one of the first monitoring frameworks by collecting multiple indicators, including 12 socioeconomic and 12 Earth system indicators. The Intergovernmental Panel on Climate Change (IPCC) Sixth Assessment Report (AR6) detailed 17 climate system components that can be attributed to human influence[3]. Butchart et al. 2010[13] listed 31 indicators related to biodiversity. The World Bank tracks over 1,500 indicators, and there are over 210 indicators for the SDGs agreed upon at the 48th session of the United Nations Statistical Commission in 2017.

The existing indicators capture both acute and chronic aspects of global change as well as socioeconomic activities. However, evaluating such a vast array of indicators is both challenging, and the information they contain often redundant[14,15]. Assessing them in isolation, without leveraging their shared or underlying information as is common in the literature[6], can lead to an underestimation of systemic risks. This highlights the need to incorporate dimensionality reduction into the monitoring and diagnosis of socioeconomic variations and global change which can better characterize synchronized or latent changes. However, simple, single-domain dimensionality reduction is insufficient because the risks associated with Earth system changes are not easily prioritized in isolation. Their full significance only emerges when their socioeconomic impacts are comprehensively considered. To date, we still lack a data-driven strategy that captures the interplay between Earth system changes and socioeconomic dynamics.

Here, we propose an interdisciplinary, data-driven paradigm that leverages interactions among multifaceted data from the biosphere, atmosphere, and socioeconomics to better understand both the anthropogenic drivers of global change and the feedback effects of environmental changes on society. This paradigm can help identify systemic biosphere-atmosphere-socioeconomic syndromes as groups of synchronized trajectories of global change and socio-economic indicators. In contrast to common empirical analyses that focus on a predefined single proxy of a socio-economic or physical variable, we advocate for interactive dimensionality reduction techniques to identify these syndromes. These multivariate approaches enhance analytical robustness by mitigating the effects of multicollinearity and accounting for variations in noise levels, thereby providing a more comprehensive and reliable representation of complex systems. We provide a prototype to identify and diagnose key socio-physical interactions using dimensionality reduction and illustrate its potential to guide climate and sustainability policies.

## UNEXPLORED INTERACTIONS BETWEEN ATMOSPHERE, BIOSPHERE, AND SOCIOECONOMICS

The atmosphere, biosphere, and socioeconomic systems interact across intra-annual, interannual, and multidecadal scales. These interactions span multiple spheres, sectors, and spatial-temporal dimensions, where high-level policies, standards, and regulations shape macroeconomic trends by influencing various microeconomic and societal sectors (Figure 1). Global change impacts the socioeconomic domain by affecting society's dependence on, and resilience to, long-term climate and environmental shifts, including extreme events. At the same time, socioeconomic activities drive changes in the atmosphere and biosphere, altering regional land use, energy systems, biosphere dynamics, and freshwater management. These shifts, in turn, contribute to planetary climate feedback, reinforcing the interconnected nature of these systems.

Socioeconomic activities that influence environmental systems can be chronic, such as greenhouse gas emissions, or acute, such as lock-down during COVID-19[16]. For example, human-induced global warming and rising atmospheric $CO_2$ levels dominate a widespread greening of terrestrial vegetation and increases in carbon sinks[17,18]. Regional vegetation greening or browning and the recent expansion of browning regimes are largely driven by the trade-off between increased temperature and land-use changes, water scarcity, or nutrient deposition or limitations[19,20]. Other chronic anthropogenic drivers such as industrialization, particularly coal-fired power generation, and deforestation due to urban sprawl and agricultural expansion, play a significant role in changing the climate so that vegetation dynamics is reshaped. However, acute interactions between societal crises and environmental changes also play a role. For example, case studies have shown that the reduction in aerosol pollution during COVID-19 moderated more favorable light conditions for European forest productivity which might promote land sinks[21,22]. On the other hand, financial or societal crises can also increase land carbon emissions as a larger share of people is pushed towards subsistence, such as increased illegal deforestation activities in Brazil[23] or agricultural expansion during COVID-19 in China[24]. Thus, socioeconomic dynamics or crises can influence atmospheric dynamics and terrestrial processes, affecting carbon or water cycling. Understanding these interactions is key to more accurately predicting future vegetation scenarios.

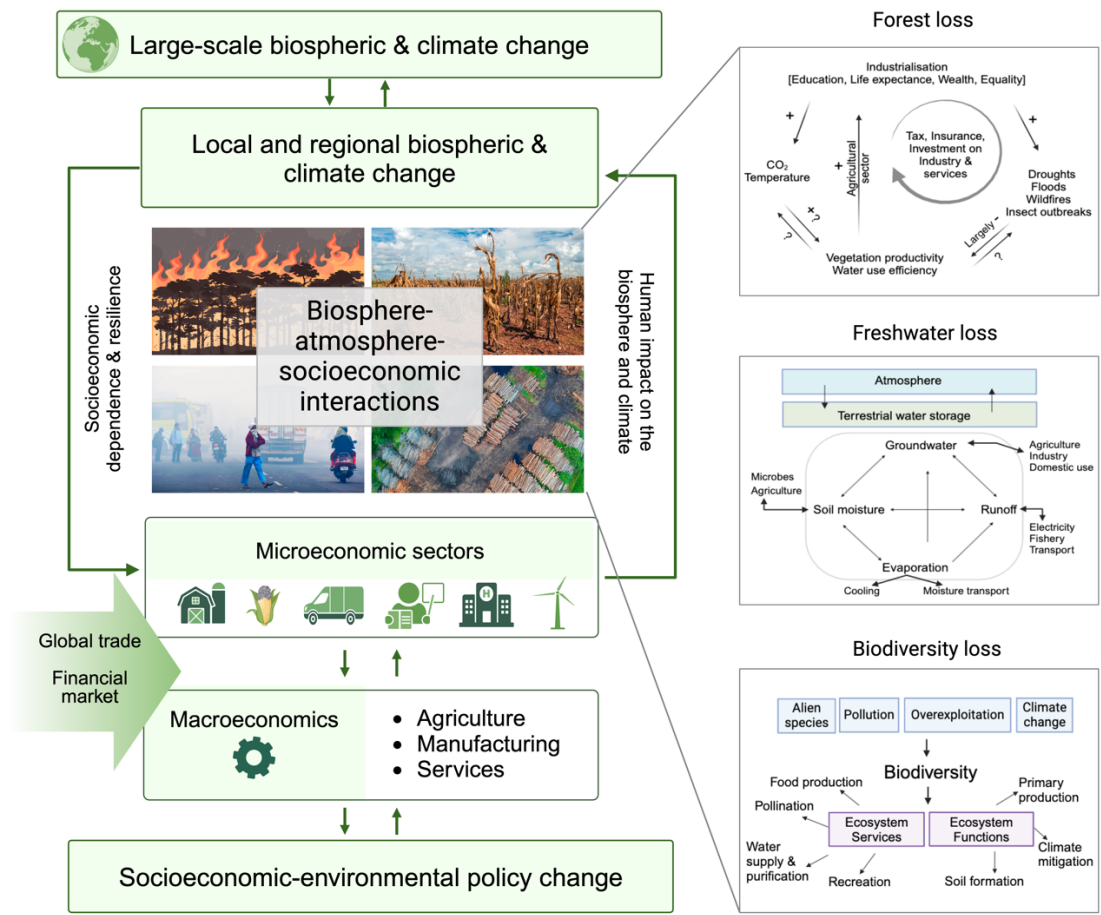

Figure 1. A schematic showing biosphere-atmosphere-socioeconomic interactions. The figure is created with BioRender.com. Inserted photos are courtesy of VORNEWS and PEXELS.

Globally, climate and biospheric extreme events are becoming increasingly intense and frequent, which can trigger complex socioeconomic and environmental interactions[25,26]. However, traditional assessments that focus solely on economic damage[27] cannot fully capture the broader loss in human well-being and environmental sustainability. For instance, the 2019–2020 Australian wildfires exemplify the intricate interactions between climate extremes and societal impacts[28]. These fires occurred alongside record-breaking heat, marked by the highest mean annual temperature, and severe drought, with the lowest annual precipitation recorded in the past 50 years[29]. The direct economic cost of these events was equivalent to more than 7% of national GDP in 2019[30]. However, indirect costs, including ecosystem recovery and community well-being, are challenging to quantify. The fires devastated approximately 20 million hectares, affected 21% of temperate forests, destroyed over 6,000 buildings, and caused significant losses of life[31]. In addition to immediate destruction, the fires emitted over 337 million tons of $CO_2$, contributing to further climate feedback loops[32]. The ecological toll was staggering, nearly a billion animals perished, and thousands of plant species, including endangered ones, were destroyed[33]. The health impacts of such wildfires extend far beyond direct fatalities. Chronic respiratory, cardiovascular, and visual impairments, as well as mental health disorders, often emerge as long-term consequences[34]. Vulnerable populations, including senior citizens, individuals with preexisting conditions, and children, are disproportionately affected. Despite burned areas of wildfires are estimated to be around 40 million km² (~4% of the global land surface) annually[35], standardized and timely indices for measuring their multifaceted impacts remains a significant knowledge gap. Furthermore, beyond wildfires, building a more comprehensive and generalized framework for monitoring and assessing impacts of various types of atmospheric or biospheric extremes is also a challenge.

Single-domain dimensionality reduction has been applied in previous studies to monitor and assess changes in the Earth system and socioeconomic conditions. Examples include constructing drought indicators[36], investigating terrestrial land surface dynamics[14], ecosystem functions[15], building biodiversity monitoring frameworks[13,38], evaluating national SDG performance[39,40], and developing multifaceted indices for human well-being[16]. However, challenges remain, particularly in developing novel approaches and practices for leveraging data-driven, interactive trajectory diagnoses across the concerned atmosphere, biosphere, and socioeconomic systems. Figure 2 illustrates major challenges with examples in the assessment workflow of various types of emerging climate and biospheric extremes, as well as of acute and chronic socioeconomic crises. These challenges include monitoring global but also regional dynamics of Earth system and socioeconomic components, integrating multiple observations into a unified system for a more efficient evaluation, and diagnosing the magnitudes and impacts of climate and biospheric extremes or long-term systemic risks. For example, Butchart et al. 2010[13] compiled indicators to track and monitor progress toward the 2010 target of the Convention on Biological Diversity, which aimed to achieve a significant reduction in the rate of biodiversity loss. They distinguished three categories of biodiversity-related indicators: the state of biodiversity, which included a wild bird index or indices from the International Union for Conservation of Nature (IUCN) Red List; the pressure, which relates to the causes of biodiversity loss; and the response, which includes measures such as protected area extent (Figure 2A). While this monitoring framework provides assessment of the causes and consequences of biodiversity loss, relying on the sum of global statistical data makes it difficult to gain insights into underlying mechanisms related to local management practices and into assessing regional yet significant biodiversity crises. In terms of integrating observations into a unified system or model, previous studies have made progress by incorporating multifaceted indices that extend beyond economic growth to include the dimension of human well-being and sustainable environment[16,39,40]. However, they do not fully account for more complete spectrum of climate and biospheric changes and their interactions with SDG performance (Figure 2B). Diagnosing Earth system changes beyond steady states, such as demonstrated in Kraemer et al. 2020[14], is valuable for post-hoc assessments of an extreme event or compound events (Figure 2C). However, the societal consequences of these events remain unintegrated in this framework, as societal consequences depend not only on the magnitude of the extreme events but also on the exposure and vulnerability of affected properties and communities[41].

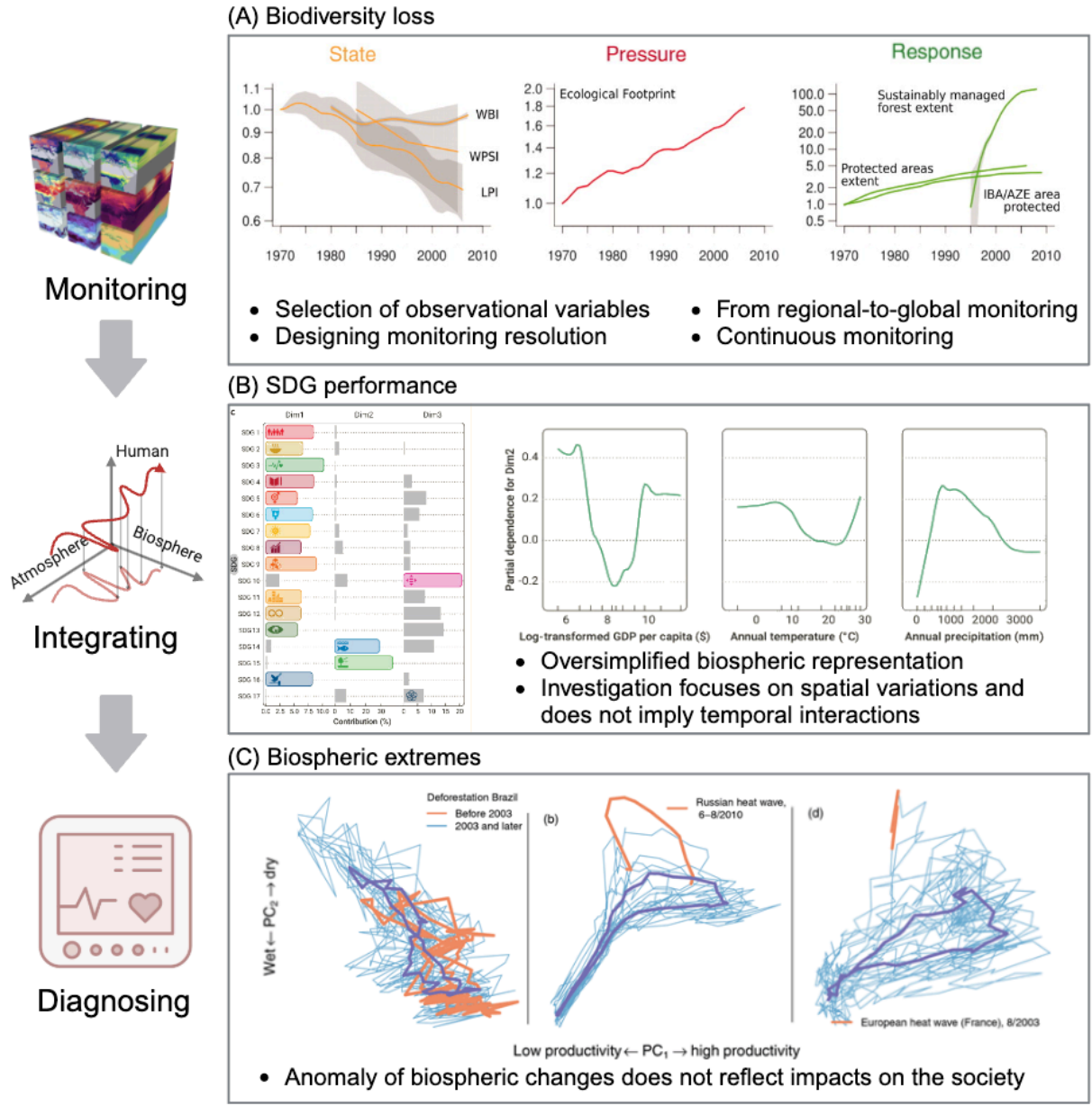

Figure 2. Challenges in monitoring, integrating, and diagnosing changes and interactions among biosphere, atmosphere, and socioeconomics. The data cube monitoring element is courtesy of Mahecha et al. 2020[42]. (A) Illustrates a case of data collection and classification of biodiversity states, pressures, and responses, adapted from Butchart et al. 2010[13]. (B) Constructing three dimensions of national SDGs by accounting for SDG variable variance across countries and illustrating the spatial dependence of the second dimension (which relates to long-term mean temperature and precipitation) on gross domestic product (GDP), temperature, and precipitation[39]. (C) Shows extreme trajectories in Brazil, Russia, and Europe by tracking major biospheric changes related to vegetation productivity and land surface wetness, courtesy of Kraemer et al. 2020[14].

## THE SYNDROME PARADIGM

As macroeconomic systems and various societal sectors interact with atmospheric and biospheric changes, a paradigm that systematically connects these interactions is beneficial for a more comprehensive understanding of socio-physical processes (Figure 1). In medicine, a syndrome is defined as a set of signs and symptoms that occur together and characterize a particular condition. By analogy, previous studies have proposed a similar concept where they predefined environmental change syndromes at the regional scale[42]. However, such approach based on predefined syndromes may overlook emerging syndromes that have not yet been conceptually established. Furthermore, the strength of identified socio-physical relevance remains uncertain.

In recent decades, the ever-growing satellite observations and in-situ measurements[44,45,46,47,48] has created a unique opportunity adopt a more flexible, data-driven approach for identifying and tracking syndromes in global change research. Newly developed satellite missions enable the monitoring of biospheric and atmospheric variables with unprecedented spectral, spatial, and temporal resolution[49,50,51]. Meanwhile, data in the recent decades can also systemically track socioeconomic changes: (i) official statistics, which cover a wide range of societal sectors[52,53,54]; (ii) web and crowdsourcing data, which provide detailed but short-term behavioral insights[55,56]; and (iii) satellite-based socioeconomic products, which allow for large-scale, high-resolution monitoring[57,58,59]. By integrating Earth system and socioeconomic data, we can more comprehensively examine interactions among the atmosphere, biosphere, and human systems (Table S1).

The complexity of socio-physical interactions, combined with significant redundancies in relevant datasets, makes manual monitoring of large-scale global datasets both costly and inefficient. To overcome this, we propose an interactive dimensionality reduction approach to integrate atmospheric, biospheric, and socioeconomic datasets. This approach allows us to capture interactions and processes across domains, project multiple variables into unified views, and identify syndromes by constructing main axes of variations through maximizing their relevance across domains, e.g., using a canonical correlation analysis[60] or its variants[61] (Figure 3A), or deep-learning algorithms to account for non-linearity across various spatio-temporal scales (Figure 3B). Low-dimensional data representations provide a proof-of-concept avenue for post-hoc assessments, including trend analysis on individual axes, extreme event detection, and socio-physical resilience quantification (Figure 3C). In some cases, temporal lag effects and teleconnections are critical for the emergence of specific syndromes and could be further explored (Figure 3D). The process of interpreting syndromes could play a crucial role in generating new hypotheses about relevant drivers and responses, which can help to inform intervention and management strategies when incorporated into a causal inference framework (Figure 3E).

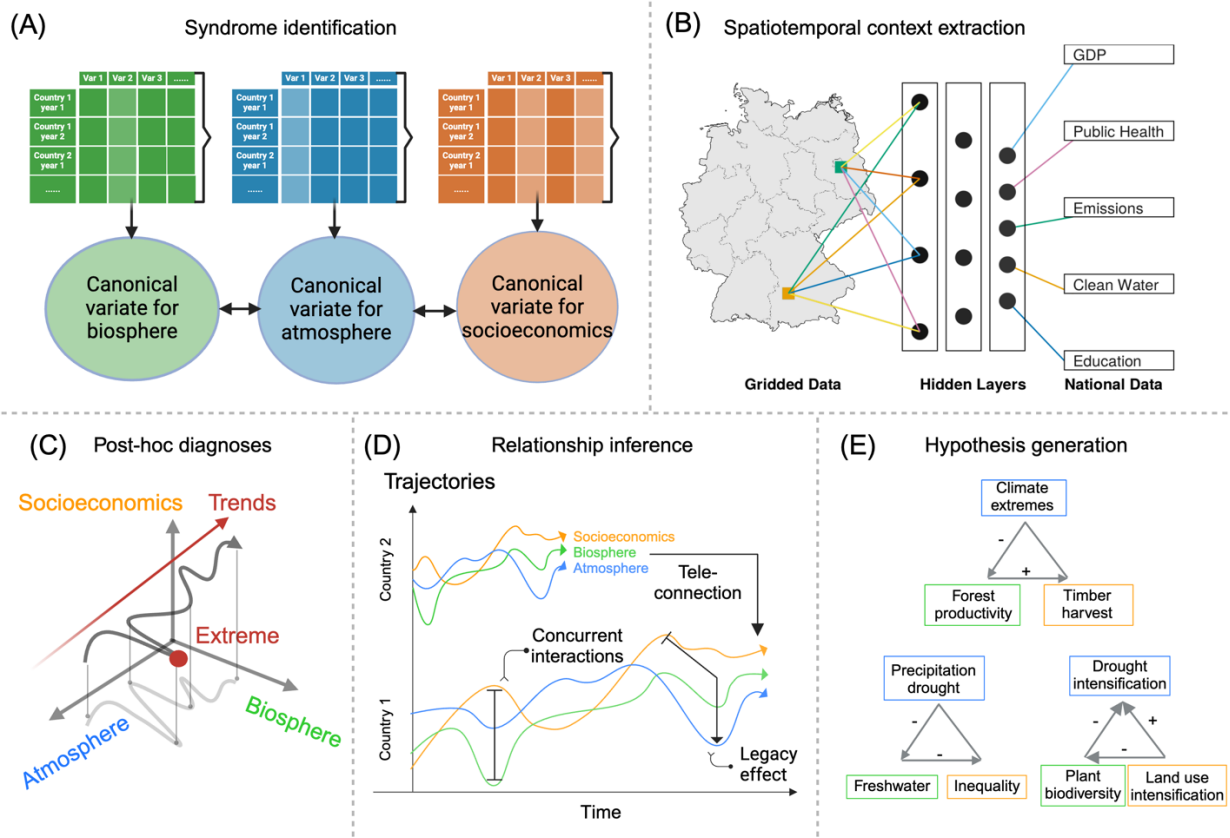

Figure 3. Data-driven syndrome illustration. (A) Canonical correlation analysis can aid in syndrome identification. Each table represents the available data in each domain, incorporating multiple variables across spatial and temporal scales. (B) Matching relevant spatio-temporal contexts using deep learning. (C) Systematic diagnosis based on integrated main axes. Each axis represents major changes within a domain that are relevant to other domains, identified through dimensionality reduction analysis. Adaptive and quantitative diagnostic approaches can be applied, such as extreme event detection or trend analysis. (D) Syndromes diagnosed based on different types of relationships: concurrent, legacy, or teleconnected relationships. (E) Identified syndromes and respective diagnoses on biosphere-atmosphere-socioeconomic interactions can inspire hypothesis testing and causal inference modeling.

### *A prototype of syndrome identification*

We propose a prototype to characterize syndromes using regularized canonical correlation analysis (CCA)[62,63] and we employ global inter-annual data at the national level during the period 2003-2022 (See details in Methods and data in Table S1). A three-way CCA method allows us to construct main axes of variation by maximizing the respective correlation between biosphere, atmosphere, and socioeconomic datasets while controlling for multicollinearity among variables within each set. Consequently, we identify syndromes based on the high relevance among cross-domain variables, extracting their synchronic changes and transformations across twenty years. Figure 4A-C shows loadings of different variables used to construct the first- and second-order CCA axes. The first axes describe integrated changes in terrestrial vegetation ecosystems, atmospheric relative humidity and radiation, and changes in agricultural and infrastructural indicators in the socioeconomic domain, so called "natural ecosystem syndrome"; the second axis integrates land surface and atmospheric changes specific to densely populated regions (inferred from variables weighted by population density and labeled as 'POP' in Figure 4B), along with changes in emissions, pollution, natural resources, governance, taxation, and economic performance—collectively referred to as the "urban syndrome." The performance of the CCA model is evaluated in Figure S1, with average pairwise correlation coefficients of approximately 0.5 for the first- and second-order CCA axes. The first- and second-order axes are relatively robust against uncollected variables in our analysis and hence we mainly interpret results from these two pairs of axes. We test the robustness in Figure S2 which presents averaged correlation coefficients of ~0.9 and ~0.5 respectively, when run multiple times of models by randomly leaving out one-fifth of the considered variables. Note that, significant trends and country-level means are removed for each variable before implementing CCA to account for any spurious correlation owing to geopolitical and historical factors.

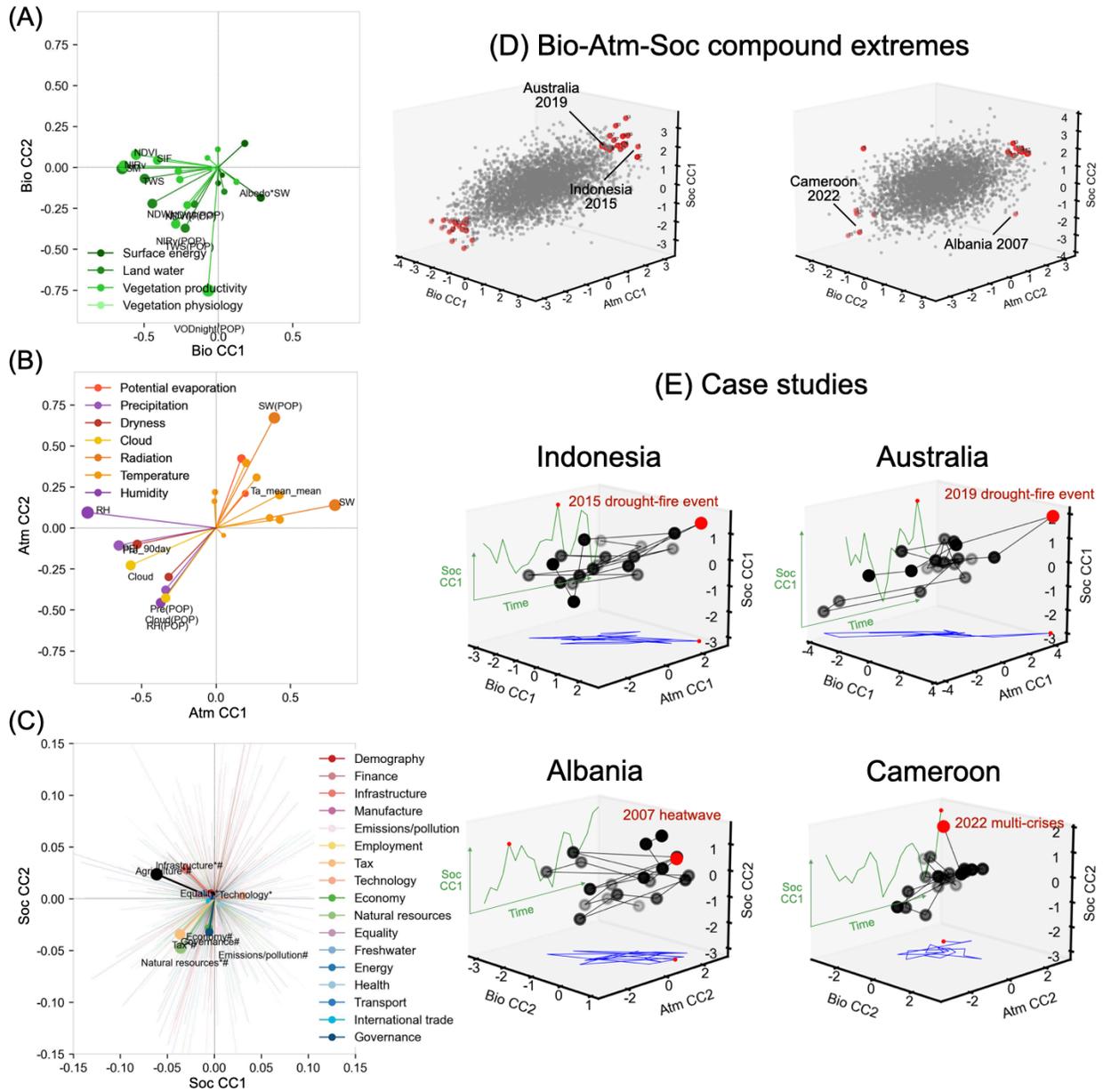

Figure 4. Main axes of temporal interactions between biosphere (Bio), atmosphere (Atm) and socioeconomics (Soc) across global countries during 2003-2022. (A-C) Biplots show the correlation coefficients between each variable and the constructed first (CC1) and second main axes (CC2) generated from Canonical Correlation Analysis (CCA). Two types of biospheric and atmospheric variables are used in the analyses in the grid data aggregation procedure: weighted average by population density (denoted by "POP") or simple average, and both are weighted by actual land areas across grid cells. Loadings in (A-C) are colored and labeled by categories of variables. Given the very high number of socioeconomic variables (over 700) in (C) (light dashed lines), the results are grouped into 17 socioeconomic categories (solid lines). The loadings from these categories are first calculated by averaging the absolute values, and the signs of the loadings are assigned by the majority sign of individual loadings. Significance of the loading signs are indicated by an asterisk (*) for x-axis or a hashtag (#) for y-axis, if more than ⅔ of the individual variables have consistent signs with the majority. By computing the Euclidean distance, we rank loadings of variables and the first half variables with high loadings are labeled in (A-C). Full labels of variables from the biosphere and atmosphere can be found in Figure S3. (D) Top-down extreme detection on the first

socioeconomic axis. Red color and numbers indicate extreme events. For example, Indonesia 2015, Australia 2019, Albania, and Cameroon are automatically detected using a threshold on the distribution of biospheric, atmospheric, and socioeconomic axes. (E) Temporal variation of CCA axes in Indonesia, Australia, Albania, and Cameroon, with lighter colors indicating earlier years and darker colors indicating later years. Data that are presented in (D-E) are normalized CCA components for each country for a fair comparison among countries. Black trajectories represent data from 3 domains, blue curves are the projected trajectories disregarding the socioeconomic axis, and green curves are the temporal trajectories of the socioeconomic axis.

The natural ecosystem syndrome is composed of multiple vegetation indices in the biospheric domain, such as the near-infrared reflectance of vegetation (NIRv)[64], the normalized difference vegetation index (NDVI)[61], sun-induced chlorophyll fluorescence (SIF)[65,66], and vegetation optical depth (VOD)[67]. Land water availability is also highly correlated with the constructed natural ecosystem syndrome, including soil moisture (SM)[45] from the ESA Climate Change Initiative and terrestrial water storage from NASA's Gravity Recovery and Climate Experiment (GRACE)[44]. All these variables are negatively correlated with the first CCA axis, with increases in the first CCA axis indicating decreases of vegetation and land water availability. This is different from variables related to the land surface energy availability, which are positively correlated with the first CCA axis. Vegetation and land water in the biosphere are positively correlated with relative humidity and precipitation in the atmosphere but negatively correlated with incoming shortwave radiation and annual mean temperature. Vegetation and land water are also positively correlated with the agricultural sector and infrastructural properties. The first-order CCA components thus highlight the central role of terrestrial vegetation ecosystems and land water availability in socio-physical processes and interactions.

The second syndrome, urban syndrome consists of variables related to land water, vegetation, and energy availability in densely populated urban regions, where positive values indicate heat and dryness stress in cities. This syndrome synthesizes changes across multiple socioeconomic sectors, suggesting that natural stress on cities is associated with and likely leads to negative consequences in various socioeconomic activities. The urban syndrome integrates measurable impacts across domains, including anthropogenic greenhouse gas emissions, particulate pollution, and natural resource extraction, as well as fiscal policy, governance effectiveness, and macroeconomic indicators (such as trade volume and new business formation). The synthesized relevance of each socioeconomic category on the constructed CCA axes is low, because each represents the mean correlation across hundreds of socioeconomic variables which include non-significant variables, and because we remove country-level means and trends in these variables to strictly control potential confounding effects. However, note that, although socioeconomic data are relatively noisy, when quantifying individual variables the highest absolute correlation for socioeconomic variables can still exceed 0.25 (Figure S3).

Various socioeconomic aspects have been widely investigated in terms of their responses to climate and land surface stress, particularly in urban areas. Urbanization intensifies emissions and pollution, with cities contributing approximately three-quarters of global $CO_2$ emissions[69]. Additionally, elevated temperatures in urban regions amplify the occurrence of extreme weather events such as droughts and wildfire[70], which in turn increase air pollution, posing significant health risks to urban populations[71,72]. Beyond direct environmental and economic consequences, extreme heat and urban heat island effects have been linked to increasing tensions and conflicts in vulnerable regions. For instance, in the Middle East and North Africa, climate change and heat-induced resource scarcity have been shown to exacerbate violent conflicts by increasing competition over water and arable land, straining governance systems, and fueling migration and instability[72]. Cities experiencing prolonged heatwaves and infrastructure stress often witness heightened social unrest, particularly in regions already grappling with political instability. For instance, the case of heat-water stress and conflict escalation in parts of Syrian, Afghanistan, Iraq, and Central America illustrates how climate-driven resource depletion can trigger disputes[74,75,76,77]. These environmental stressors also have profound economic consequences, reducing labor or livestock productivity and increasing costs in retail, leading to overall economic downturns[78]. This interconnected set of challenges highlights the role of environmental stress in shaping urban socioeconomic conditions, reinforcing the necessity of integrated policies that address climate resilience and sustainable urban development.

*Applications on diagnosing socio-physical extremes*

Figure 4D demonstrates that utilizing syndromic approaches enables the diagnosis of influential extreme events through a top-down methodology. For simplicity, we implement a percentile threshold-based (<5th or > 95th for each axis) method to detect impactful biosphere-atmosphere-socioeconomic compound extremes across various countries and years. This approach diagnoses extreme events and evaluates them by not only examining abnormal environmental stressors but, more critically, highlighting crises or downturns in socioeconomic performance. Illustrative cases include the 2015 Indonesian drought and wildfires, the 2019 Australian drought and fires, the 2007 food scarcity crisis in Albania, and the 2022 multifaceted crises in Cameroon (Figure 4E). We note that the extreme detection approach we apply is simplified for identifying case studies rather than for prediction or quantifying the return rate of extreme events. Future studies focusing exclusively on extreme events should incorporate the joint probability distribution of the multivariate variables or employ alternative threshold criteria[79].

The top-down identification of these extreme events aligns closely with event-specific evaluations and holds the potential to uncover previously unrecorded losses. For instance, the 2015 Indonesian drought and fires, driven by strong El Niño conditions (see temporal trajectories of some biospheric and atmospheric variables in Supplementary Figure S4), resulted in severe environmental degradation, widespread haze, and economic losses estimated at US$28 billion, underscoring the complex interplay between climate extremes and land-use stress[80]. Similarly, the 2019 Australian drought and fires exemplify the cascading effects of extreme heat, vegetation stress, and socioeconomic disruptions (see temporal trajectories of some individual socioeconomic variables in Supplementary Figure S5). In 2007, Albania has experienced some energy shortages and urban heat stress during the European heatwave which partly demonstrate its socioeconomic instability when confronted with climatic stressors. In 2022, Cameroon experienced a complex crisis driven by post-COVID-19 economic instability[81] and ongoing regional conflicts of Anglophone Cities of Cameroon[82], while also facing climate-related disasters. Severe flooding in the Far North region affected nearly 40,000 individuals, destroying homes and agricultural lands[83]. These case studies illustrate that the syndromic approach strengthen the characterization of compound risks in vulnerable regions resulting from multiple stressors where some events were not extensively analyzed in previous literature.

### *Potential limitations and essential criteria*
Certain limitations of the syndrome paradigm must be acknowledged. First, understanding the relevant scales remains challenging because local and global syndromes capture different dynamics, and transferring macro-level insights into actionable local policies requires further refinement (Box 1). while a global assessment—such as that of the 2022 multifaceted crises in Cameroon (Figure 4E)—might indicate broad trends related to coastal flooding risks, it cannot accurately predict accurate impacts of rising sea levels on urban infrastructure and community livelihoods or forecast future flood losses in major coastal cities[84,85]. To address this gap, fine-tuned syndrome analyses that leverage granular theme-specific socioeconomic data are necessary. Moreover, global and national-level syndrome analyses typically rely on averaged atmospheric and biospheric data, which provide only a preliminary understanding and likely fail to capture nuanced impacts on local socioeconomic variations. Besides, recent advances in deep learning have demonstrated promising capabilities for bridging the gap between global and local analyses. For example, recent studies have developed foundation models for the Earth system which has successfully integrated diverse, multi-resolution data across various spatio-temporal scales[86]. Although this approach requires a deluge of training data, it overcomes traditional limitations in matching coarse global datasets with fine-grained local or sectoral socioeconomic activities. Integrating deep learning-based dimensionality reduction methods give the potential to identify localized, theme-specific syndromes, ultimately supporting more effective policy and decision-making.

Second, caution is warranted when handling variables that are strongly physically linked. For example, if highly correlated data from the biosphere and atmosphere are treated as if they belong to separate domains, the insights derived may predominantly reflect their inherent correlation rather than an equal incorporation of socio-physical interactions. An example of this is the relationship between land surface temperature (LST) and air temperature. In this case, rather than using LST directly, we employ a ratio between LST and air temperature, which largely rules out direct temperature impact and explicitly incorporate ecosystem functional changes into the biospheric domain in the framework. Moreover, although fine spatio-temporal data is desirable, the use of merged multi-stream data, through downscaling or reconstruction algorithms, must be approached with care. This is because these data-driven products may inadvertently carry dependent information from their input sources, such as meteorological variations, potentially obscuring novel insights in a syndrome analysis.

Additionally, dimensionality reduction models must balance between the interpretability of a model and the representation of socio-natural non-linearity, ensuring that results are not overfitted or misleading. Lastly, while syndromes help diagnose climate extremes, attributing causality remains complex, necessitating prudent interpretation to avoid overgeneralization or oversimplification. Overall, effective implementation of the syndromes concept outlined here requires information-rich multi-scale observations across sectors, and a detailed analytical framework, both of which we expect to advance significantly as the approach becomes more widely adopted.

*Box 1. Essential criteria for a data-driven syndrome paradigm.*

Criteria:

(1) Comprehensively understanding relevant scales:

A better understanding of syndromes, from local to global, from short term to long term, are all milestones towards a more sustainable future. Long-term syndrome studies or global syndrome studies emphasize overviews of global change and macroeconomics, while locally tailored syndrome studies reveal more changes in microeconomic sectors, which can inform regulation by local authorities.

(2) Data criteria:

Production: Fine spatio-temporal resolution is essential.

Selection and pre-processing: Strategies related to efficient and sustainable big-data storage and processing; Strategies of tailoring data depend on the number of data sampling and the level of data independence.

(3) Model selection of dimensionality reduction:

A balance between considering the nature of socio-natural non-linearity in the model selection and a convincing socio-biophysical interpretation.

(4) Interpretation of syndromes:

Not over-interpret causality directly on the syndrome results, even when embedding it with a causal model. A prudent attitude is needed, considering limitation of certain models and data uncertainties.

(5) Uncertain reduction:

Dependent datasets putting into different domains should increase biases.

Constructed trajectories shall be reproduced by a major subset of data (see Figure S2).

Country-level mean values and variable trends likely introduce spurious correlation and need to be removed.

## OUTLOOK

The syndrome paradigm offers a data-driven approach to monitoring, integrating, and diagnosing the dynamic interactions between the biosphere, atmosphere, and socioeconomic systems. It aims to bridge multidisciplinary knowledge and data, integrating joint environment-society trajectories to uncover both existing and emerging mechanisms related to anthropogenic impacts on the Earth system, as well as the socioeconomic stability and resilience to global environmental changes. By employing dimensionality reduction techniques, we identify and classify highly interactive changes across biospheric, atmospheric, and socioeconomic components, which can be leveraged for risk assessment and enhancing early warning capabilities. We present a prototype framework that defines the natural ecosystem and urban syndromes, summarizing key interactions within these syndromes. The syndrome analysis can help identify influential extreme events for an improvement of disaster preparedness and resilience planning, particularly in climate-vulnerable nations or nations with monolithic economic structure. Despite its advantages, implementing the syndrome paradigm requires addressing challenges in data integration, variable selection, and model interpretation. Future research must embrace a multidimensional perspective to integrate data-driven assessment frameworks to guide policies for a more sustainable and adaptive future.

## METHODS

We apply regularized Canonical Correlation Analysis (CCA) across three domains—biosphere, atmosphere, and socioeconomics—to develop a data-driven syndrome prototype. This approach integrates interactive changes across these domains into a few principal axes by maximizing their correlation[62,63]. The three-way regularized CCA is defined as:

$$\max \left( w_1^T X_1^T X_2 w_2 + w_2^T X_2^T X_3 w_3 + w_3^T X_3^T X_1 w_1 \right), \text{ subject to:}$$

$$(1 - c_1) w_1^T X_1^T X_1 w_1 + c_1 w_1^T w_1 = 1,$$

$$(1 - c_2) w_2^T X_2^T X_2 w_2 + c_2 w_2^T w_2 = 1,$$

$$(1 - c_3) w_3^T X_3^T X_3 w_3 + c_3 w_3^T w_3 = 1.$$

Where $c_1, c_2, c_3$ denote the regularization parameters for each of the three domains. $w_1, w_2, w_3$ denote canonical weight vectors for each domain. $w_1^T, w_2^T, w_3^T$ denote transpose of the canonical weight vectors. $X_1, X_2, X_3$ denote data matrices for each domain. And $X_1^T, X_2^T, X_3^T$ denote transpose of the data matrices. The max function maximizes the pairwise cross-covariances of the three domains' canonical variates, thereby capturing their shared variability.

We apply the CCA model globally at inter-annual and national scales (2003–2022), using biosphere, atmosphere, and socioeconomic data, which are listed in Table S1 under the label 'used'. Variables were selected based on four criteria: they cover the period 2003–2022, have global coverage at national and inter-annual scales (or finer spatio-temporal resolutions), remain independent across domains to avoid strong physical dependencies, and be interpretable in terms of their physical or socioeconomic meaning.

The socioeconomic data were preprocessed from the World Development Indicators (WDI) database, which contains over 1400 variables. First, we selected approximately 1000 variables, removing those that were strongly dependent on country size (e.g., total population or total agricultural area) while retaining proportion-based metrics. Next, countries or variables with more than 70% missing data were removed. To fill remaining data gaps, we applied Probabilistic Principal Component Analysis (PPCA)[88], which explained over 90% of the variance in the socioeconomic domain. The robustness of the resulting dimensions was tested across multiple runs.

For the biospheric and atmospheric data, we aggregated gridded data to a national level using both normal averaging and population-weighted averaging, and using weights based on actual land areas for each grid cell. Population-weighted averages highlight biospheric and atmospheric variability in densely populated regions, such as urban areas and coastlines. Two versions of those variables were used in training the CCA model. Averaged differences between countries and long-term trends in variables are influenced by geographical and historical geopolitical factors, while temporal interactions between nature and society reflect real-time environmental changes. To explicitly study concurrent inter-annual changes across the three domains, we remove mean values at the country level and remove significant long-term trends.

The performance of the CCA model is evaluated in Figure S1, where the average pairwise correlation coefficients are approximately 0.6 for the first-order CCA axis and 0.4 for the second-order CCA axis (Figure S1b). Each variate of CCA axes has relatively low explained variance which indicates that only a small proportion of data variation is relevant for the other two domains at the national level (Figure S1c-g). The socioeconomic variates have the lowest explained variance which are related to the relatively larger noise level. The first pair of CCA axes is the most robust, even when accounting for potential omitted variables, followed by the second pair of CCA axes (Figure S2). As illustrated in Figure S2, randomly removing 20% of all variables still reconstructs similar CCA axes in biosphere, atmosphere, and socioeconomics, respectively. This demonstrates the robustness of our approach in identifying highly interactive changes across these domains, which can be further analyzed for the global compound event detection.

## RESOURCE AVAILABILITY

### Lead contact

Further information and requests for resources should be directed to and will be fulfilled by the lead contact, Wantong Li (wantongli@berkeley.edu).

### Materials availability

This study did not generate new unique materials.

### Data and code availability

The codes required for reproducing the results and figures in the main text have been deposited at https://doi.org/10.5281/zenodo.14996785, as well as the data to run the codes are available at https://doi.org/10.5281/zenodo.14876723.

## ACKNOWLEDGMENTS

The authors thank Daniel E. Pabon Moreno, Weijie Zhang, Daniel Loos, and Lazaro Alonso Silva at the Max Planck Institute for Biogeochemistry for fruitful discussions and technical supports. W.L. was supported by the Open-Earth-Monitor Cyberinfrastructure project and she received funding from the European Union's Horizon Europe research and innovation programme under grant agreement No. 101059548. M.R. and G.D. acknowledge funding by the European Research Council (ERC) Synergy Grant 'Understanding and modelling the Earth System with Machine Learning (USMILE)' under the Horizon 2020 research and innovation programme (Grant Agreement No. 855187). F.G. and M.D.M. acknowledge funding by the European Space Agency (ESA) through the DEEPESDL Project.

## AUTHOR CONTRIBUTIONS

Conceptualization, M.R., G.D., and F.G.; methodology, W.L., M.R., G.D., F.G., G.K., and M.D.M.; investigation, W.L.; writing—original draft, W.L.; writing—review & editing, W.L., M.R., G.D., G.K, D.F., A.C., T.K., J.S., F.G., M.M., M.D.M. and U.W.; funding acquisition, M.R., G.D., and F.G.; resources, W.L., F.G., J.S., and U.W.; supervision, M.R. and G.D.

## DECLARATION OF INTERESTS

The authors declare no competing interests.

## SUPPLEMENTAL INFORMATION

Supplementary information including Table S1 and Figures S1-5.